\documentclass[twocolumn]{aastex631}

\usepackage{amsmath}
\usepackage{amssymb}
\usepackage{graphicx}
\usepackage{booktabs}
\usepackage{multirow}
\usepackage{xcolor}
\usepackage{hyperref}
\usepackage{needspace}

\newcommand{\jwst}{{JWST}}

\renewcommand{\sharp}{\hbox{\sc sharp}}

\begin{document}

\title{Tracing the Luminosity Evolution of Globular Cluster Systems with JWST: MACS J1149.5+2223 at $z=0.544$}

\author[0000-0001-6265-0541]{Jessica M. Berkheimer}
\affiliation{School of Earth and Space Exploration, Arizona State University, Tempe, AZ 85287, USA}

\author[0000-0001-8156-6281]{Rogier A. Windhorst}
\affiliation{School of Earth and Space Exploration, Arizona State University, Tempe, AZ 85287, USA}

\author[0000-0001-8762-5772]{William E. Harris} 
\affiliation{Department of Physics \& Astronomy, McMaster University, Hamilton ON L8S 4M1, Canada}

\author[0000-0001-6650-2853]{Timothy Carleton}
\affiliation{School of Earth and Space Exploration, Arizona State University, Tempe, AZ 85287, USA}

\author[0000-0002-6150-833X]{Rafael Ortiz III}
\affiliation{School of Earth and Space Exploration, Arizona State University, Tempe, AZ 85287, USA}
\email{rortizii@asu.edu}

\author[0000-0002-9895-5758]{S.\ P.\ Willner}
\affiliation{Center for Astrophysics \textbar\ Harvard \& Smithsonian,
60 Garden Street, Cambridge, MA, 02138, USA}


\correspondingauthor{Jessica M. Berkheimer}
\email{jberkhei@asu.edu}

\begin{abstract}
We present a study of the compact-source population associated with the massive galaxy cluster MACS~J1149.5+2223 at $z=0.544$ using deep \jwst/NIRCam imaging in F115W, F150W, and F200W. After photometric and morphological selection, we identify 1,202 unresolved compact-source candidates. The observed luminosity distribution samples only the bright side of the expected globular-cluster luminosity function (GCLF), remaining several magnitudes above the turnover. Based on published GCLF turnovers for Abell~2744 and MACS~0417, we adopt an expected turnover magnitude of $m_0=33.66$ mag for MACS~1149. A forward-modeled fit then yields $\sigma=1.71$ mag, which characterizes the observed bright-side slope under the adopted turnover rather than the intrinsic width of the full GCLF\null. Comparison with PARSEC simple stellar population models shows that the elevated luminosities of the compact sources are consistent with the expected fading of younger stellar populations as they evolve toward the present epoch. The brightest sources reach luminosities comparable to those of ultra-compact dwarfs. However, stellar-population degeneracies and possible foreground contamination make it difficult to distinguish. Together with recent \jwst\ studies of globular cluster systems across a range of redshifts, MACS~1149 extends the emerging observational sequence of compact stellar systems and supports stellar population fading as an important contributor to their luminosity evolution across cosmic time.
\end{abstract}



\section{Introduction}

The James Webb Space Telescope (\jwst) has opened a new observational window on globular cluster systems (GCSs) beyond the local Universe. Owing to their great ages and long dynamical survival times, globular clusters (GCs) preserve important information about the earliest phases of galaxy formation and assembly, making their integrated properties powerful tracers of galaxy evolution \citep[e.g.,][]{Brodie2006,Forbes2018,Beasley2020}. \jwst's unprecedented combination of angular resolution and near-infrared sensitivity now makes it possible to identify and characterize individual GCs in galaxy clusters at intermediate redshifts, where the observed populations represent earlier stages in their evolutionary history. GCSs are not static populations, but continue to evolve through stellar evolution and dynamical processes over cosmic time \citep{Fall2001,Vesperini2001,Kruijssen2015}. Observations at intermediate redshift therefore provide a direct opportunity to test these evolutionary predictions by observing GCSs at earlier stages in their development than those found in the nearby Universe.

Recent \jwst\ observations have begun extending detailed studies of GCSs to massive galaxy clusters at cosmological distances, including Abell~2744, MACS~J0416.1$-$2403, MACS~J0417.5$-$1154, RXJ~2129.7+0005, and El Gordo \citep{Harris2025A2744I,Harris2025A2744II,Harris2025MACS0417,Harris2025ElGordo,keatley2025jwst,berkheimer2026globular}. Collectively, these studies are providing the first direct view of GCSs at progressively earlier stages in their evolution. 

\begin{figure*}[t]
    \centering
    \includegraphics[width=\textwidth]{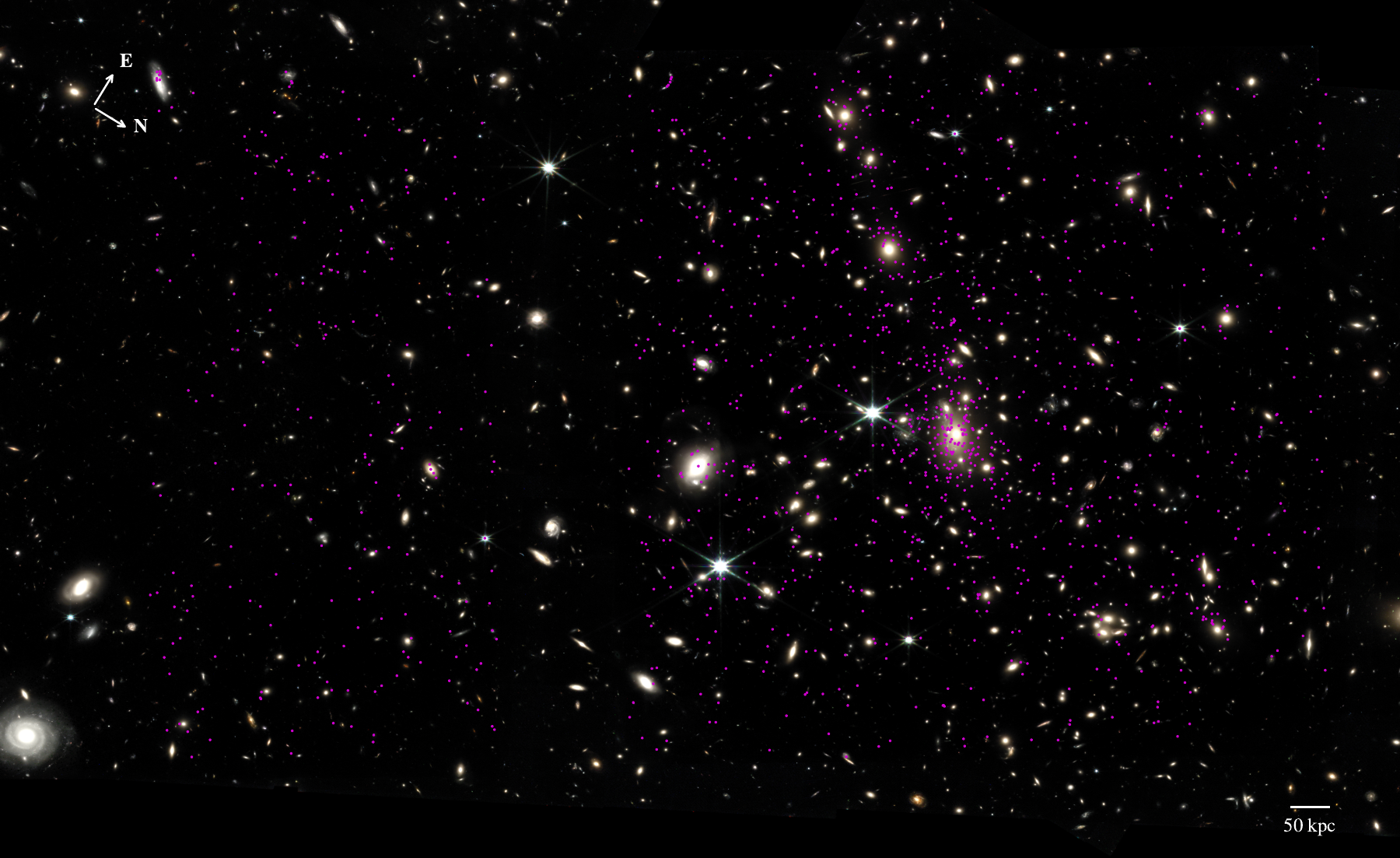}
    \caption{
    Three-color \jwst\ NIRCam image of the MACS~1149 field constructed from the F115W (blue), F150W (green), and F200W (red) mosaics. Magenta points indicate the locations of the 1,202 sharpness-selected unresolved compact-source candidates used in this work. The orientation compass indicates North and East, and the scale bar corresponds to 50 kpc at the cluster redshift ($z=0.544$).
    }
    \label{fig:macs1149_rgb}
\end{figure*}

MACS~J1149.5+2223 (hereafter MACS~1149) is a massive galaxy cluster at redshift $z=0.544$, corresponding to a look-back time of approximately 5.5~Gyr. As one of the six \emph{Hubble} Frontier Fields clusters, MACS~1149 has been the subject of extensive multiwavelength observations and benefits from exceptionally deep imaging \citep{Ebeling2007,Postman2012,Lotz2017}. The cluster core has a dominant brightest cluster galaxy, while the larger-scale system contains multiple substructures with its ongoing merger \citep{Smith2009}. The cluster contains a rich population of massive elliptical galaxies embedded within an extensive intracluster environment, making it an excellent laboratory for investigating GCSs in a dense cluster environment at intermediate redshift. As such, MACS~1149 provides an important addition to the growing sample of \jwst\ studies tracing the evolution of GCSs across cosmic time.

GCSs are expected to evolve photometrically over time as their stellar populations age and their constituent clusters gradually lose mass through stellar evolution and long-term dynamical processes \citep{Fall2001,Vesperini2001,Kruijssen2015}. Together, these processes produce predictable changes in the integrated luminosities, colors, and LFs of GC populations. Consequently, GCSs observed at intermediate redshifts are expected to have clusters that are both more luminous at a fixed mass and, having experienced less mass loss, are more massive than their counterparts in the local Universe.

In this paper, we investigate the GCS of MACS~1149 using deep, multi-epoch \jwst/NIRCam imaging obtained in the F115W, F150W, and F200W filters. The resulting photometric properties are interpreted within the context of recent \jwst\ studies and simple stellar population models to assess the evolutionary state of the system. Throughout this paper we adopt the Planck 2016 cosmological parameters, $H_0 = 67.8~\mathrm{km~s^{-1}~Mpc^{-1}}$, $\Omega_{\rm M}=0.308$, and $\Omega_\Lambda=0.692$ \citep{Planck2016}. At the cluster redshift of $z=0.544$, this cosmology gives
a luminosity distance  $d_{\rm L}=3216$~Mpc corresponding to a distance modulus $(m-M)_0=42.54$ mag. The angular-diameter distance is $d_{\rm A}=1349$~Mpc giving a physical scale of $6.54~{\rm kpc~arcsec^{-1}}$. The Galactic foreground extinction from NED is $A_V=0.063$ mag \citep{SchlaflyFinkbeiner2011}.
Unless otherwise stated, all magnitudes presented in this work are in the AB system.

Section \ref{sec:obs} describes the observations and data reduction, while Section \ref{sec:photometry} presents the photometric measurements and completeness analysis used to construct the globular cluster candidate catalog. Section~\ref{sec:gcs} characterizes the observed photometric properties of the globular cluster system, including its color--magnitude distribution, LF, and brightest compact objects. These observations are interpreted using simple stellar population models in Section~\ref{sec:ssp} to investigate the evolutionary state of the MACS~1149 GCS and place it within the emerging observational sequence of intermediate-redshift GCSs. Finally, we summarize our results and discuss their implications for the photometric evolution of GCSs.

\section{Observations and Data Reduction}
\label{sec:obs}
The \jwst/NIRCam observations of MACS~1149 used in this work were obtained as part of the PEARLS GTO Program 1176 (PI: R.~Windhorst; \citealt{Windhorst2023}).
The reduced NIRCam mosaics used were produced by the SKYSURF-IR project, a JWST Cycle 3 Archival Legacy program designed to characterize the near-infrared sky background using existing JWST/NIRCam observations along approximately 100 independent sightlines (PI: R.~Windhorst). The mosaics are sampled at a pixel scale of 0\farcs03 pixel$^{-1}$ and encompass the central region of MACS~1149. Although the full mosaics have large angular extent, the analysis was restricted to the central region of the stacked images, where the overlapping observations provide the deepest and most uniform exposure coverage. This region therefore offers the highest photometric sensitivity and the most reliable completeness characterization while encompassing the majority of the globular-cluster-candidate population considered in this work.

The photometric analysis was performed on the F115W, F150W, and F200W science mosaics, which have intensity units of MJy~$\mathrm{sr}^{-1}$. Prior to source detection and point-spread function (PSF) photometry, the cropped images were linearly scaled by a factor of $10^4$ and offset by 1000 counts to facilitate processing within the DAOPHOT package \citep{Stetson1987} in IRAF\null. This transformation preserves the relative photometric information while ensuring that all pixel values remain positive throughout the analysis.

To maximize the sensitivity to faint compact sources, the three science mosaics were combined to form a deep multi-filter detection image. An initial source catalog was generated from this stacked image with a detection threshold of approximately $4\sigma$, where $\sigma$ is the standard deviation above the local background noise. The detection image was used exclusively for source identification, while DAOphot PSF-fitting photometry was subsequently performed individually on the F115W, F150W, and F200W mosaics.

\section{Photometry and Completeness Tests}
\label{sec:photometry}
\subsection{Photometric Measurements}

Empirical PSFs were constructed independently for the F115W, F150W, and F200W images using bright, isolated, unsaturated stars distributed throughout the science field. Because GCs at the redshift of MACS~1149 are effectively unresolved in \jwst\ imaging, accurate characterization of the PSF is essential for precision photometry \citep{Stetson1987}. These PSF models were then used to perform PSF-fitting photometry (fitting both position and magnitude) for all detected sources in each filter. Measured fluxes were converted to calibrated AB magnitudes following the {\jwst}/NIRCam photometric calibration. Aperture corrections were derived from isolated, unsaturated point sources by comparing the PSF-fitting magnitudes with large-aperture measurements obtained within a radius of 10 pixels = 0\farcs30. The resulting correction, $\Delta m_{10}=m_{10{\rm px}}-m_{\rm PSF}$, was determined independently for each filter. An additional correction from the 10-pixel aperture to infinite radius was then applied using the JWST/NIRCam encircled-energy curves \citep{NIRCamHandbook}. At \(r=0\farcs30\), approximately 88\% of the total point-source flux is enclosed; an infinite-aperture correction of $\Delta m_{\infty}\simeq-0.145$ mag was adopted. The calibrated magnitudes were computed as
\begin{equation}
m_{\rm AB}
=
-2.5\log_{10}\left(\frac{f}{10~{\rm nJy}}\right)
+28.90,
\end{equation}
where $f$ is the aperture-corrected flux in nJy. The $K$-corrections used for comparisons with stellar-population models are described in Section~5. 

\begin{figure}[ht]
\centering
\includegraphics[width=0.5\textwidth]{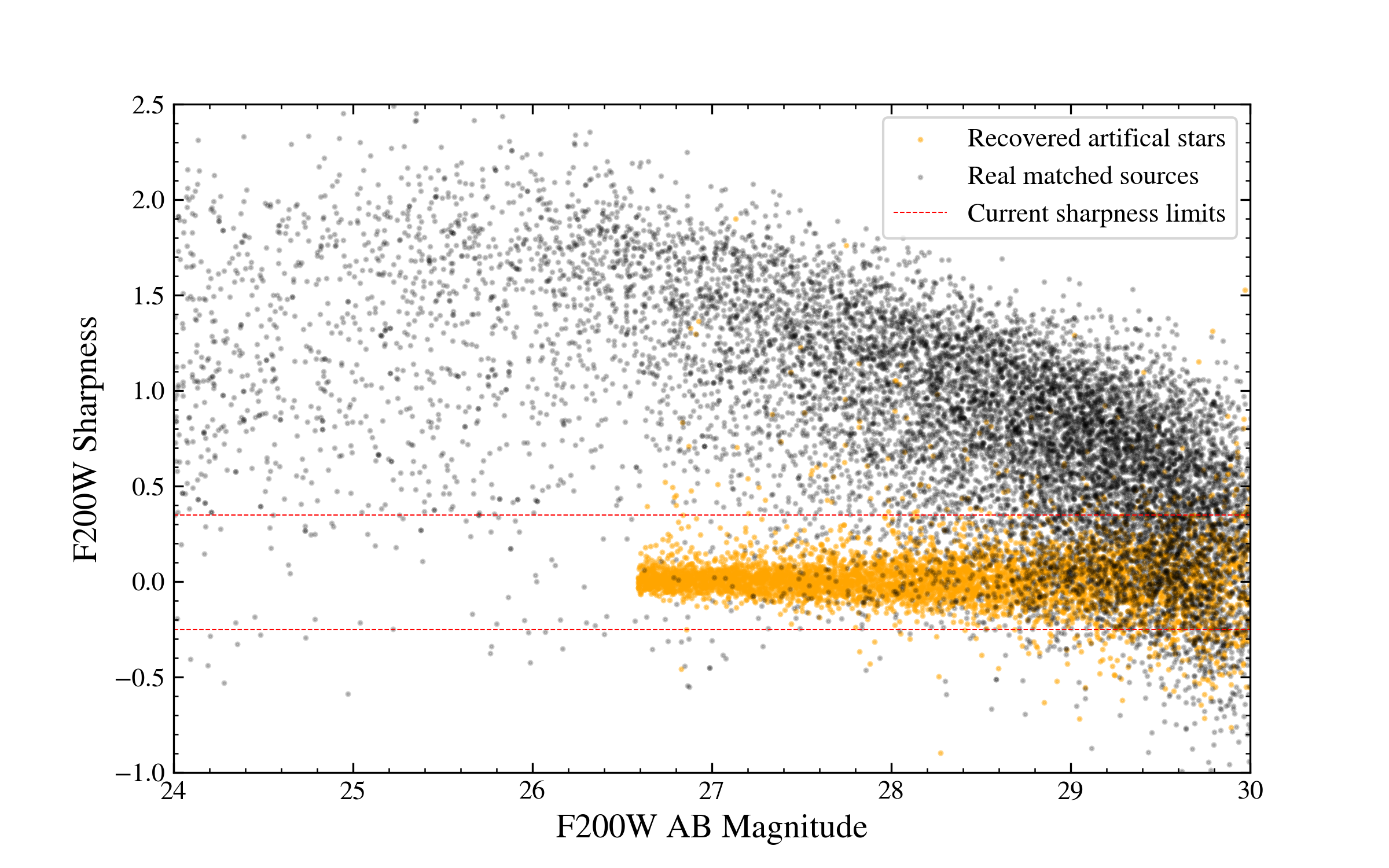}
\caption{
DAOPHOT \sharp\ as a function of F200W magnitude for the science catalog and recovered artificial stars. The dashed lines indicate the adopted \sharp\ selection limits. Because the F200W artificial-star distribution exhibited the broadest \sharp\ distribution among the three filters, the same limits were applied independently in all three filters.
}
\label{fig:sharp}
\end{figure}

The resulting F115W, F150W, and F200W catalogs were cross-matched using a positional tolerance of 2 pixels, and sources were required to have been detected and measured in all three filters. Morphological selection was performed using the DAOPHOT \sharp\ parameter as shown in Figure \ref{fig:sharp}. Based on the artificial-star tests described in Section~3.2, candidate sources were required to satisfy $-0.25 \leq \hbox{\sharp} \leq 0.35$ independently in F115W, F150W, and F200W. Objects outside this interval were rejected as likely extended sources, unresolved blends, or image artifacts. Figure~\ref{fig:sharp} illustrates the adopted {\sharp} selection in F200W\null. Following the application of these selection criteria, the final catalog contains 1,202 unresolved compact-source candidates. This catalog defines the parent sample, with completeness-limited subsets adopted for the subsequent quantitative analyses where appropriate.

\subsection{Artificial-Star Tests}

The completeness of the GC candidate catalog was quantified through artificial-star tests performed with the DAOPHOT \texttt{ADDSTAR} task. Artificial stars generated from the empirical PSF in each filter were
inserted into the science images in 30 independent realizations of 300 stars each, yielding a total of 9,000 injected sources. The stars were distributed across the primary cluster field and spanned the magnitude range relevant to the analysis. Limiting the number of injected stars in each realization prevented the simulations from artificially increasing the level of crowding.

Each artificial-star realization was processed using  source-detection and PSF-fitting procedures identical to those applied to the science images. Recovered measurements were associated with the injected sources using a 2-pixel positional matching radius. For each injected star, the recovered
position, magnitude, and photometric properties were recorded whenever a valid match was obtained.

We distinguish between positional recovery and recovery into the final science catalog. A source was considered positionally recovered when valid measurements were obtained in all three filters. To be counted as recovered in the final completeness analysis, the source was additionally required to satisfy the same \sharp\ selection applied to the science catalog ($-0.25 \leq \sharp \leq 0.35$ in each filter). Of the 9,000 injected artificial stars, 5,380 were positionally recovered in all three filters,
while 4,956 remained after application of the \sharp\ selection. Thus, 424 sources (7.9\% of the three-filter positional recoveries) were excluded by the morphological criteria.

\begin{figure}[ht]
\centering
\includegraphics[width=0.5\textwidth]{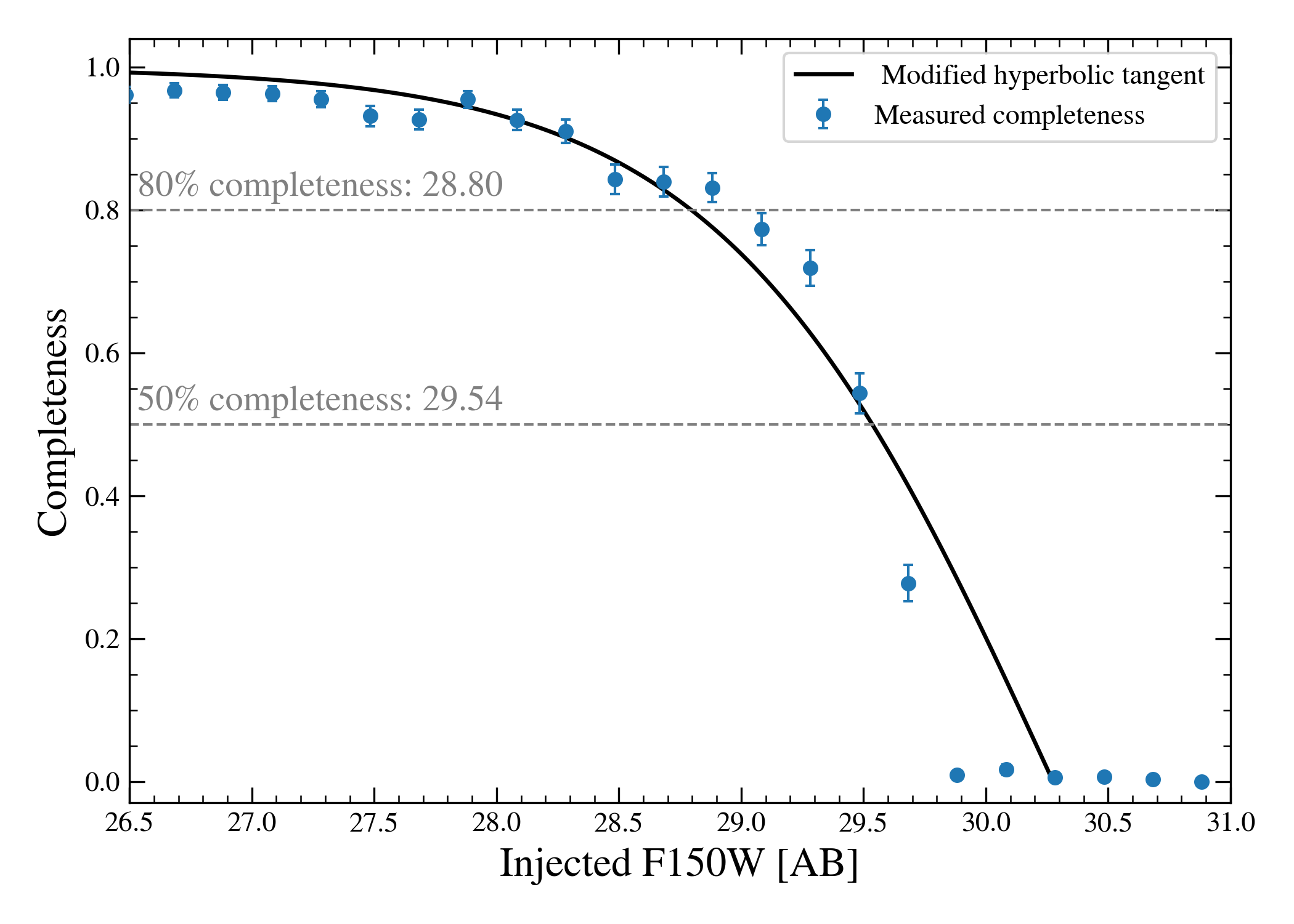}
\caption{
Artificial-star completeness as a function of injected F150W magnitude. Points show the measured completeness in each magnitude bin, while error bars represent the corresponding binomial uncertainties. The solid curve shows the adopted completeness function, and the dashed lines indicate the 80\% and 50\% completeness levels at $m_{\rm F150W}=28.80$ and 29.54 mag.
}
\label{fig:completeness}
\end{figure}

\begin{figure}[ht]
\centering
\includegraphics[width=0.45\textwidth]{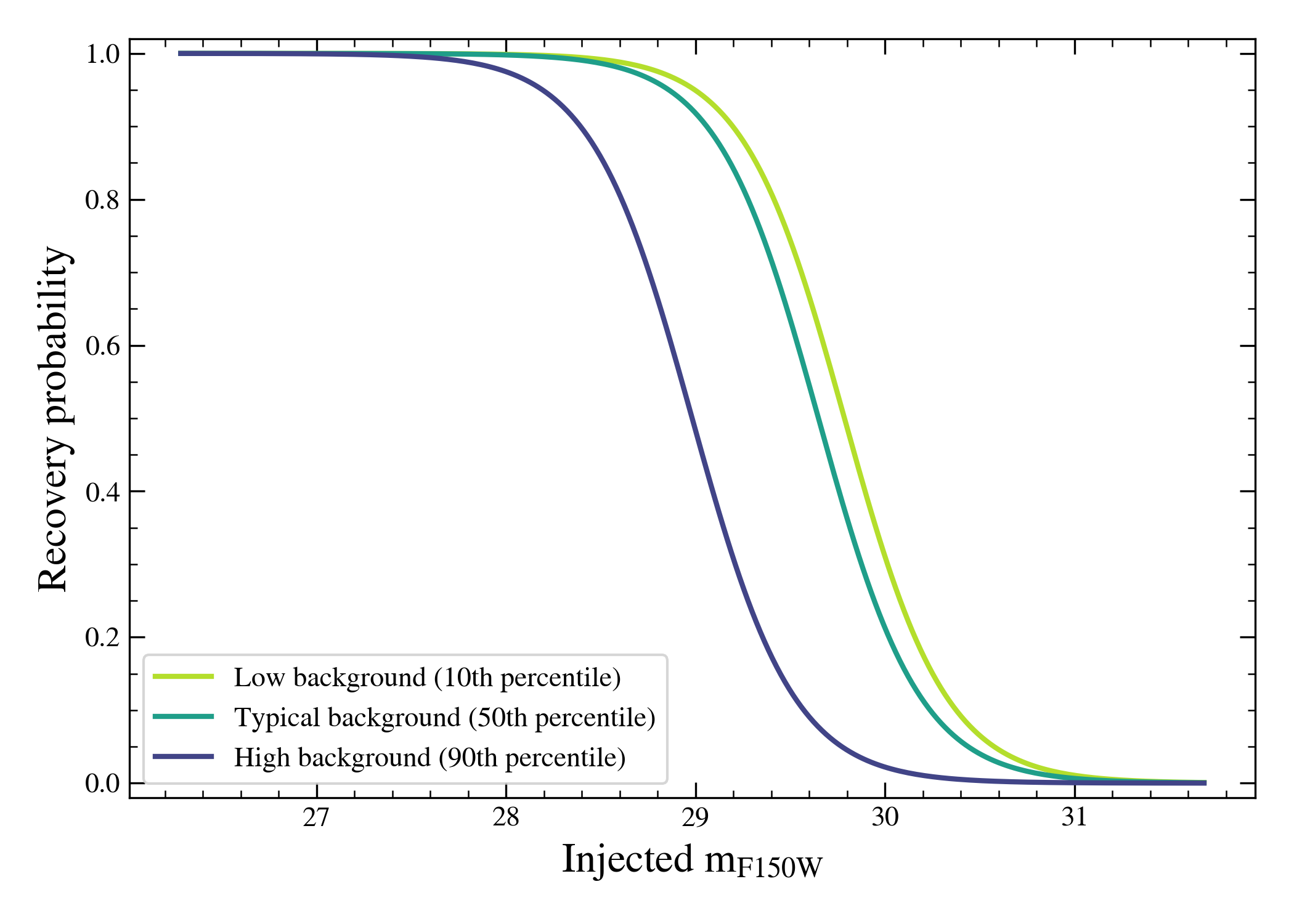}
\caption{Predicted artificial-star recovery probability as a function of injected F150W magnitude for representative local-background levels. Curves show predictions from the logistic-regression model evaluated at the 10th, 50th, and 90th percentiles of the ${\rm STDEV}_{150}$ distribution, corresponding to low, typical, and high local backgrounds, respectively. The reduced recovery efficiency at the highest background levels illustrates the increased incompleteness within the bright cores and halos of cluster galaxies.}

\label{fig:environment_completeness}
\end{figure}

Completeness was calculated in 0.2-mag bins of injected F150W magnitude as the fraction of artificial stars satisfying the final three-filter recovery criterion. The uncertainty in each bin was computed assuming binomial counting statistics,
\begin{equation}
\sigma_p=\sqrt{\frac{p(1-p)}{N}},
\end{equation}
where $p$ is the measured completeness in the bin and $N$ is the total number of injected artificial stars within that magnitude bin. To provide a continuous representation of the recovery efficiency, the
binned completeness measurements were fit with the modified hyperbolic tangent function used by \citet{Harris2025A2744I},
\begin{equation}
f(m)=
\frac{2}{1+\exp[\beta(m-m_1)]}-1,
\end{equation}
where $\beta$ controls the rate of the completeness decline and $m_1$ is the magnitude at which this transformed function crosses zero. The fit was restricted to the well-sampled portion of the
completeness relation (\(f>0.20\)); the fitted function was not used beyond \(m_1\), where it would become nonphysical as a completeness fraction. The best-fitting parameters are $\beta=1.49$ and $m_1=30.27$. Figure~\ref{fig:completeness} shows the measured completeness together with the best-fitting function. Solving the fitted relation yields 80\% and 50\% completeness limits of $m_{\rm F150W}=28.80$ and $29.54$ mag, respectively. These limits are given on the aperture-corrected, observed-frame F150W magnitude scale used for the artificial star tests. The magnitude-dependent completeness relation was subsequently used to characterize the recovery efficiency of the final GC candidate sample.

The structured intracluster light and bright galaxy halos in MACS~1149 may produce spatial variations in source recovery that are not captured by a purely magnitude-dependent completeness function. We therefore performed an additional diagnostic analysis using logistic regression to model the binary artificial-star recovery outcome as a function of injected F150W magnitude and the local background dispersion ${\rm STDEV}_{150}$, measured within the DAOPHOT sky annulus. The recovery probability was modeled as
\begin{equation}
    \mathrm{logit}(p)=\beta_0 + \beta_1 m_{\rm F150W}
    + \beta_2 \log_{10}({\rm STDEV}_{150}+1)
\end{equation}
where $p$ is the probability that an artificial star satisfies the final three-filter recovery criteria \citet{Harris2025A2744II,Harris2025MACS0417}. The fitted coefficients are $\beta_1=-3.73\pm0.10$ and $\beta_2=-3.17\pm0.15$, where the uncertainties are the $1\sigma$ standard errors from the logistic regression. The resulting model achieved a classification accuracy of 95.2\% in 10-fold cross-validation. The contribution of the local-background term was assessed by comparison with an otherwise identical magnitude-only model using a likelihood-ratio test. Including the background term significantly improves the fit ($\Delta\chi^2=615$ for one additional degree of freedom; $p=1.1\times10^{-135}$), showing that recovery depends not only on source brightness but also on the local background.

Figure~\ref{fig:environment_completeness} shows the recovery probability predicted by the two-variable model as a function of injected magnitude evaluated at the 10th, 50th, and 90th percentiles of the local background distribution. These curves represent model predictions at representative background levels rather than independently measured completeness relations.  The low- and median-background predictions are nearly indistinguishable, whereas only the highest-background regions show a noticeable reduction in recovery efficiency. Thus, although local background contributes significantly to the recovery probability, its practical effect is most apparent in the bright galaxy core and halos of cluster galaxies. The relatively weak background dependence across the majority of the field supports our use of the global magnitude-dependent completeness relation for the GC candidate sample.


\begin{figure*}[t]
\centering
\includegraphics[width=\textwidth]{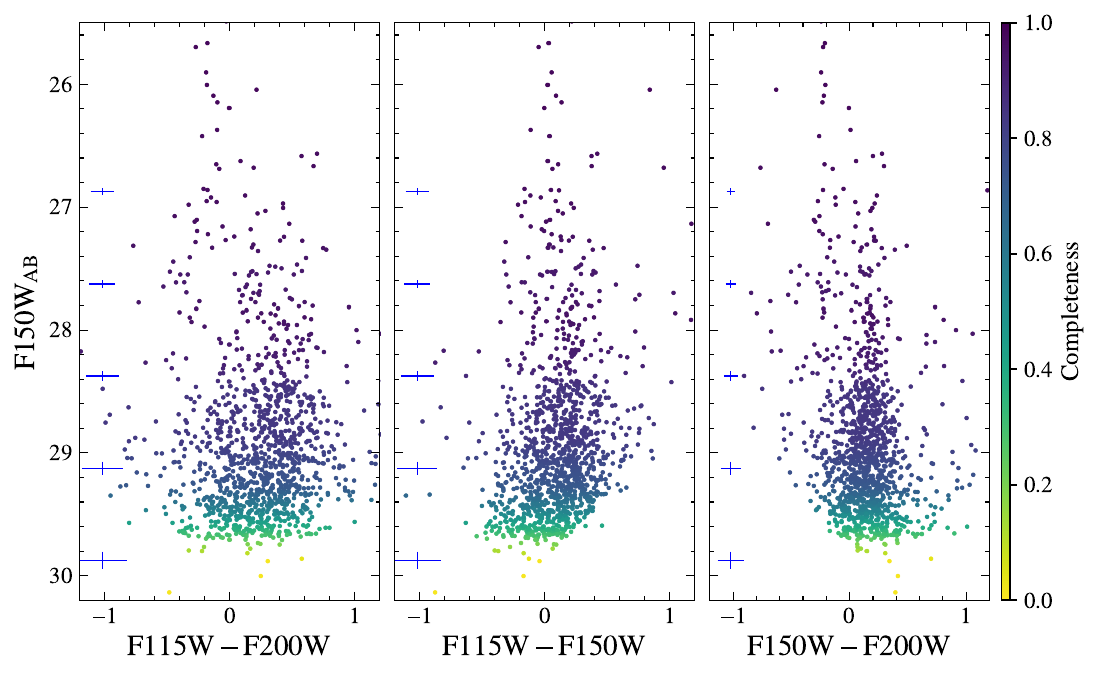}
\caption{Color--magnitude diagrams of the final GC candidate sample described in Section~\ref{sec:photometry}. The panels show the $(\mathrm{F115W}-\mathrm{F200W})$, $(\mathrm{F115W}-\mathrm{F150W})$, and $(\mathrm{F150W}-\mathrm{F200W})$ colors as a function of F150W magnitude. Each point is colored according to its completeness, computed from the magnitude-dependent completeness function derived in Section~3.2 (Figure~\ref{fig:completeness}). Representative photometric uncertainties are shown at the left of each panel and were computed from the median magnitude errors in bins of F150W magnitude.}
\label{fig:cmd}
\end{figure*}

\section{Properties of the GC System}
\label{sec:gcs}

Having established the photometric selection and completeness of the compact-source catalog, we now turn to the observed properties of the MACS~1149 population. We first examine its distribution in color--magnitude space and then characterize the luminosity function of the sources associated with the main cluster region.  
\subsection{Color--Magnitude Diagram}
The final GC-candidate sample of 1,202 objects is shown in Figure~\ref{fig:cmd} as three color--magnitude diagrams (CMDs). 
As expected, the uncertainties increase toward fainter magnitudes, with colors involving F115W exhibiting larger uncertainties than F150W$-$F200W because of the larger photometric uncertainties in the F115W measurements. 
As demonstrated in recent studies \citep{Harris2025A2744I,Harris2025MACS0417,berkheimer2026globular}, the $(\mathrm{F150W}-\mathrm{F200W})$ color index is less sensitive to GC metallicity than the other color combinations and therefore exhibits the narrowest intrinsic color sequence. The modest increase in scatter toward fainter magnitudes is consequently attributable primarily to increasing photometric uncertainties. This comparatively narrow color baseline provides a useful representation of the compact-source population and is adopted for the stellar-population comparisons presented in Section~\ref{sec:ssp}.

\subsection{Globular Cluster Luminosity Function}

The globular cluster luminosity function (GCLF) is one of the primary observational diagnostics of GCs. It provides important constraints on both their initial formation and their subsequent evolution through stellar and dynamical processes \citep{Harris2001,Brodie2006,Jordan2007,Villegas2010}. In present day GCSs, the GCLF is observed to have an approximately lognormal form, characterized by a well-defined turnover, with a declining number of clusters towards both the bright and faint ends \citep{Harris2001,Brodie2006,Jordan2007}. Its present-day shape reflects the combined effects of the cluster initial mass function, stellar population evolution, and long-term dynamical processes that preferentially disrupt low-mass clusters \citep{Fall2001,Vesperini2001,Kruijssen2015}. In nearby galaxy clusters, sufficiently deep observations routinely sample the GCLF beyond its turnover, allowing its shape and normalization to be measured directly. At intermediate redshifts, however, current \jwst\ observations generally reach only the bright portion of the LF \citep{Harris2025A2744I,Harris2025A2744II,Harris2025MACS0417,Harris2025ElGordo,berkheimer2026globular}. Although the turnover remains beyond the observational limits of these datasets, the accessible bright end of the GCLF still provides valuable constraints on the photometric evolution of globular cluster systems and enables meaningful comparisons among the growing sample of intermediate-redshift clusters. 
The GCLF is conventionally described by a Gaussian parameterization \citep[e.g.,][]{Harris2001,Jordan2007},
\begin{equation}
N(m) = A \exp\left[-\frac{(m-m_0)^2}{2\sigma^2}\right],
\label{eq:gclf}
\end{equation}
where $m_0$ denotes the turnover magnitude, $\sigma$ the dispersion, and $A$ the normalization of the LF\null. In sufficiently deep observations that extend beyond the turnover, these parameters may be determined simultaneously from the observed LF\null. However, when only the bright portion of the GCLF is sampled, 
the observed LF no longer contains sufficient information to uniquely distinguish between  $m_0$ and  $\sigma$.

\begin{figure}[ht]
\centering
\includegraphics[width=0.45\textwidth]{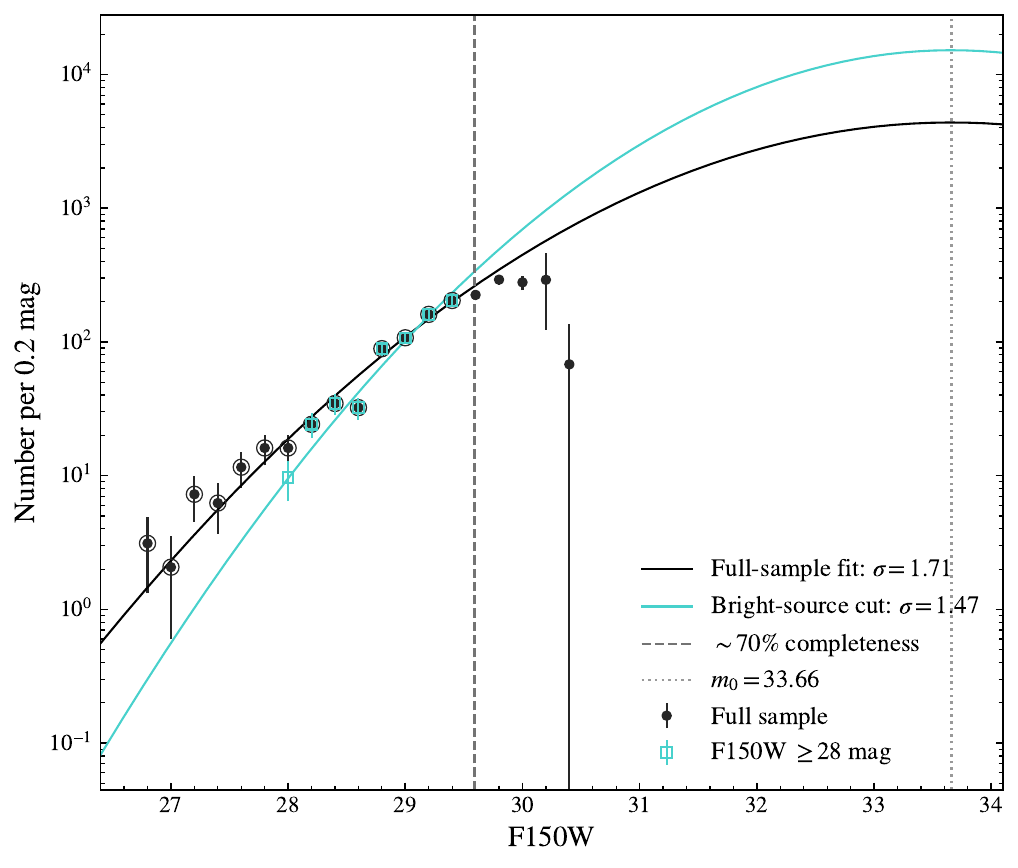}
\caption{F150W luminosity function for 983 unresolved compact sources in the primary cluster region of MACS~1149. Open circles indicate bins with recovery fractions $f_{\rm comp}>0.70$ used in the fit. The solid black curve shows the best-fitting GCLF with the turnover fixed at $m_0=33.66$ mag, yielding a bright-side dispersion of $\sigma=1.71$. The solid cyan line shows the best-fitting GCLF with the same restrictions, but also excluding the brightest points which may be contaminated by UCDs, yielding a narrower $\sigma=1.47$. The dashed line marks the 70\% completeness limit at F150W $=29.59$.}
\label{fig:gclf}
\end{figure}

Figure~\ref{fig:gclf} presents the completeness-corrected GCLF of the MACS~1149 GC candidates surrounding the primary galaxy cluster concentration. The correction was applied by dividing the observed counts in each magnitude bin by the corresponding
recovery fraction measured from the artificial-star tests. For this analysis, we restrict the sample to the NIRCam module containing the primary cluster concentration (the right-hand side of Figure~\ref{fig:macs1149_rgb}), excluding the spatially separated population in the opposite module. This selection leaves 983 sources and provides a well-defined sample associated with the main cluster environment without incorporating the more extended candidate population into the GCLF normalization. 
As shown in Figure~\ref{fig:gclf}, the observed LF rises steadily towards fainter magnitudes over the range reliably sampled by our observations but does not approach the expected turnover at $m_0=33.66$~mag. This behavior confirms that the present data sample only the bright portion of the underlying LF, consistent with recent \jwst\ studies of intermediate-redshift GCSs. Because the turnover is not observed, the shape of the bright-side LF can be reproduced by different combinations of $m_0$ and $\sigma$, producing a degeneracy between these two parameters \citep{HanesWhittaker1987,Harris2014}.

To assess this degeneracy in the MACS~1149 data, we initially allowed both the turnover magnitude $m_0$ and the dispersion $\sigma$ to vary freely. Rather than converging to a unique solution, the fits followed a clear trend. As the turnover was shifted to fainter magnitudes, the dispersion increased, allowing the models to reproduce the observed bright end of the LF with statistically similar fits. This behavior demonstrates that the present observations do not uniquely constrain both $m_0$ and $\sigma$ simultaneously, consistent with recent \jwst\ analyses of intermediate-redshift GCSs where the turnover likewise remains beyond the observational limits of the available data \citep{Harris2025MACS0417}.

Because of the inability to separate $m_0$ and  $\sigma$, an external constraint on the turnover magnitude is required. Rather than treating $m_0$ as a free parameter, we estimated its expected value by comparing MACS~1149 with the two intermediate-redshift galaxy clusters whose GCLF turnovers have been characterized most robustly to date, Abell~2744 and MACS~0417 \citep{Harris2025A2744I,Harris2025MACS0417}. Assuming that MACS~1149 follows the empirical trend established by recent \jwst\ studies of intermediate-redshift GCSs, the published turnover magnitudes were shifted by the difference in distance modulus between each reference cluster and MACS~1149. This procedure yields predicted turnover magnitudes of $m_0=33.65$ and 33.67 mag, respectively. The excellent agreement between these independent estimates indicates that the expected turnover magnitude for MACS~1149 is insensitive to the choice of the reference cluster. 
We therefore adopted $m_0=33.66$ mag as the expected turnover for the MACS~1149 GCLF. 

The range of GCLF dispersions measured in nearby and intermediate redshift systems provides useful context for interpreting the MACS~1149 LF. In nearby massive galaxies, GCLF dispersions are typically $\sigma=1.2$--1.4 and increase modestly with host-galaxy luminosity \citep{Jordan2007,Villegas2010,Harris2014}. Recent \jwst\ observations suggest that the bright-end LF of less dynamically evolved systems may be broader; in MACS~0417, for example, \citet{Harris2025MACS0417} found that $\sigma=1.6$ provides a good representation of the observed bright side. The possibility of broader GCLFs at earlier evolutionary stages is also qualitatively consistent with dynamical models, which predict that the depletion of lower-mass clusters becomes progressively more important as GCSs evolve \citep{ReinaCampos2022}. Although the present observations do not reach the faint side of the GCLF where the evolution is expected to be most apparent, these measurements and theoretical expectations provide an important reference for interpreting the bright-end dispersions of MACS~1149. 

We fit the observed bright-side LF with the turnover fixed at $m_0=33.66$ mag, allowing the dispersion $\sigma$ and normalization $A$ to vary. We restricted the fit to the magnitude bins with empirical three-filter artificial-star recovery fractions greater than 70\%, after transforming the artificial-star magnitudes to the same aperture and $K$-corrected magnitude systems as the science catalog. The faintest bin satisfying this criterion is centered at F150W $=29.4$ mag. The completeness-corrected counts shown in Figure~\ref{fig:gclf} were obtained by dividing the observed number of sources in each bin by its corresponding recovery fraction. For the fit itself, we retained the observed integer counts and forward-folded the Gaussian model through the measured completeness function, thereby accounting for the expected loss of sources in each magnitude bin. Maximizing the Poisson likelihood yields a best-fitting bright-side dispersion $\sigma=1.71$ mag. 

Because the turnover itself is not directly observed, the fitted value of $\sigma$ should not be interpreted as a measurement of the full intrinsic GCLF width. Instead, it represents the dispersion required by the adopted Gaussian parameterization to reproduce the slope of the observed bright-side LF for the assumed turnover magnitude. To assess the extent to which the most luminous compact sources influence this result, we repeated the fit after excluding sources brighter than $m_{\rm F150W}\simeq28.0$ mag, corresponding to the luminosity of a $5\times10^7\,M_\odot$, 3~Gyr, $[\mathrm{m/H}]=-2.0$ PARSEC SSP. The cut removes 67 of the 983 sources and reduces the fitted dispersion from $\sigma=1.71$ to $1.47$. As shown in Figure \ref{fig:gclf}, the difference is concentrated at the luminous end of the LF, while the two distributions converge at fainter magnitudes. This suggests that an additional population of luminous compact sources, potentially including UCDs or stripped nuclei, could be contributing to the broad bright-side LF measured. A similar excess in the bright end has been observed in other galaxy clusters, including MACS~0416, G165, Abell~2744 \citep{Harris2025A2744I,Harris2025A2744II,berkheimer2026globular,Hinrichs2026}. Although these populations vary among systems, their recurrence suggests that the brightest compact sources may not all belong to a single, underlying GC luminosity distribution.

The bright-side dispersion inferred for MACS~1149 ($\sigma=1.71$) is broader than the $\sigma\simeq1.6$ found for MACS~0417 and the typical $\sigma\simeq1.2$--$1.4$ measured for GCLFs in nearby massive galaxies \citep{Jordan2007,Villegas2010}. This comparison should be interpreted with caution, however, because the MACS~1149 value is conditional on the adopted turnover magnitude. Stellar-population evolution may also affect the location of the GCLF turnover. If the younger populations in MACS~1149 produce a brighter turnover than the adopted $m_0=33.66$, the inferred value of $\sigma$ would change accordingly. Because the turnover is not observed, the present data cannot distinguish between these possibilities.  Nevertheless, the bright-side LF and its sensitivity to the most luminous sources may be consistent with emerging evidence that the high-luminosity portions of the GCLFs at intermediate redshift differ from those of dynamically evolved systems in the nearby Universe. Whether this difference represents evolution in the global GCLF shape or changes confined mostly to the high-luminosity population cannot be determined from the present observations.

 
\section{Stellar Population Models}
\label{sec:ssp}

\begin{figure}[t]
\centering
\includegraphics[width=0.5\textwidth]{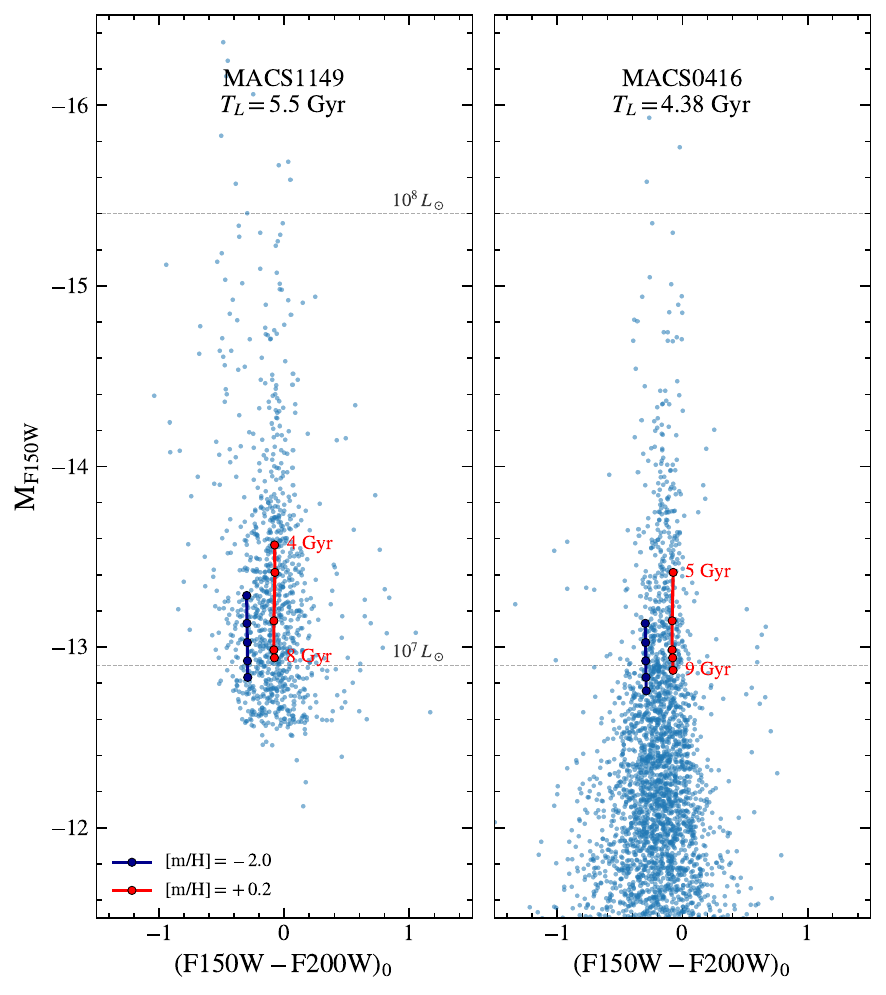}
\caption{Rest-frame CMDs for the compact-source populations in MACS~1149 (left), and MACS~0416 (right); where $(\mathrm{F150W-F200W})_0$ denotes the rest-frame color. The photometry includes the filter-dependent $K$-corrections appropriate for each system. PARSEC SSP sequences are shown for a fixed stellar mass of $2\times10^7\,M_\odot$ and metallicities of $[\mathrm{m/H}]=-2.0$ (blue) and $+0.2$ (red), with model ages appropriate for the lookback time $T_L$ of each system. Horizontal dashed lines mark the $10^7$ and $10^8L_\odot$ F150W luminosity reference levels described in the text. 
}
\label{fig:ssp}
\end{figure}

The luminosity distribution examined through the GCLF provides only one view of the evolving compact-source population. Additional information is encoded in the integrated colors of the clusters themselves, which reflect the ages and chemical compositions of their underlying stellar populations \citep{Worthey1994,BruzualCharlot2003,Maraston2005}. GCs are particularly well suited to this type of analysis because the majority of their constituent stars formed over relatively short time intervals and share broadly similar chemical compositions. Their integrated photometric properties can therefore be modeled to first order as simple stellar populations (SSPs) characterized by a common age and metallicity \citep{Brodie2006}. As these populations age, the progressive evolution of their most luminous stars causes their integrated luminosities to fade and their mass-to-light ratios to increase, while the broadband colors depend on both stellar age and metallicity \citep{BruzualCharlot2003,Maraston2005,Conroy2013}. Consequently, the location of a GC in color--magnitude space provides a useful means of examining the photometric evolution of its underlying stellar population. 

The evolutionary perspective becomes particularly important for GCSs observed at intermediate redshifts, where clusters are seen several billion years earlier in their evolution than comparable systems in the nearby Universe. At these epochs, younger stellar populations are expected to be more luminous at fixed stellar mass
\citep{BruzualCharlot2003,Maraston2005,Conroy2013}. The growing sample of GCSs observed with \jwst\ now provides an opportunity to examine this evolution across cosmic time. Recent comparisons of massive galaxy clusters spanning a range of lookback times reveal a systematic decline in the luminosity of their GC/UCD sequences toward the present epoch, closely following the behavior expected from the age-fading of simple stellar populations \citep{Harris2025ElGordo,Harris2025MACS0417}. In this context, the luminous compact-source population observed in MACS~1149 provides another snapshot of GCS evolution at an intermediate epoch. Observed elevated luminosities cannot be interpreted as higher stellar mass alone: at the epoch of MACS~1149, younger stellar populations produce substantially more light per unit mass than their present-day descendants.

To place MACS~1149 within the emerging observational picture,  Figure~\ref{fig:ssp} compares the intrinsic color--magnitude distribution with that of the GC population in MACS~0416 at $z=0.397$ \citep{berkheimer2026globular}, observed approximately 1 Gyr later in cosmic time. Following the framework of \citet{Harris2025ElGordo,Harris2025MACS0417}, PARSEC SSP models \citep{bressan2012} provide a common reference for examining how the luminosities and color of massive cluster populations are expected to evolve over this interval. The photometry for both systems is transformed to the rest frame using filter-dependent $K$-corrections calculated with the \textsc{rescuer} star-cluster $K$-correction tool \citep{ReinaCamposHarris2024}. We retain the rest-frame F150W and F200W system used in recent JWST GCS studies of \citet{Harris2025ElGordo,Harris2025MACS0417}, allowing MACS~1149 to be compared directly with these systems on the same color--magnitude scale.

For reference, Figure~\ref{fig:ssp} also marks F150W luminosity levels of $10^7$ and $10^8\,L_\odot$. We adopt an absolute solar magnitude of $M_{{\rm F150W}, \odot}=4.59$ AB mag for the JWST/NIRCam F150W band \citep{Willmer2018}. The corresponding absolute magnitude at a given F150W luminosity is then
\begin{equation}
    M_{\rm F150W}=
    M_{{\rm F150W},\odot}-2.5\log_{10}
    \left(\frac{L_{\rm F150W}}{L_{\odot}}\right).
\end{equation}
Thus, $10^7\,L_\odot$ corresponds to $M_{\rm F150W}=-12.91$, while $10^8\,L_\odot$ corresponds to $M_{\rm F150W}=-15.41$ mag. These fixed luminosity levels are intended as reference points for comparing the GC populations and SSP sequences between the two epochs and not for defining boundaries between GCs and UCDs.

The SSP sequences in Figure~\ref{fig:ssp} illustrate the expected evolution of massive GCs over the interval separating the two systems. At fixed stellar mass and metallicity, the models fade with increasing age while exhibiting relatively little evolution in $(\rm F150W-F200W)_0$. This behavior results in the nearly vertical model sequence seen in both panels and reflects the weak age sensitivity of this near-infrared color for evolved populations. The horizontal separation between the two models is driven mainly by metallicity, with the metal-poor models bluer than metal-rich ones. Between the epochs represented by MACS~1149 and MACS~0416, the shift towards older model ages therefore appears as a decrease in F150W luminosity rather than a substantial change in color.

The observed MACS~1149 population is broadly consistent with the simple evolutionary picture. Its GC/UCD candidates occupy a wide range in luminosity, extending from a sparse population approaching $10^8\,L_\odot$ to the survey's sensitivity limit. The $2\times10^7\,L_\odot$ SSP sequences pass through the main body of the observed distributions, showing that the high luminosities of many of these sources are consistent with massive stellar systems observed at an early evolutionary stage. The comparison with MACS~0416 places MACS~1149 within the broader redshift sequence identified by \citet{Harris2025MACS0417}. At the same time, the brightest MACS~1149 sources remain substantially more luminous than the fixed-mass SSP sequences. 

\subsection{Stellar-Mass Dependence of the SSP Models}
\begin{figure}[ht]
    \centering
    \includegraphics[width=0.5\textwidth]{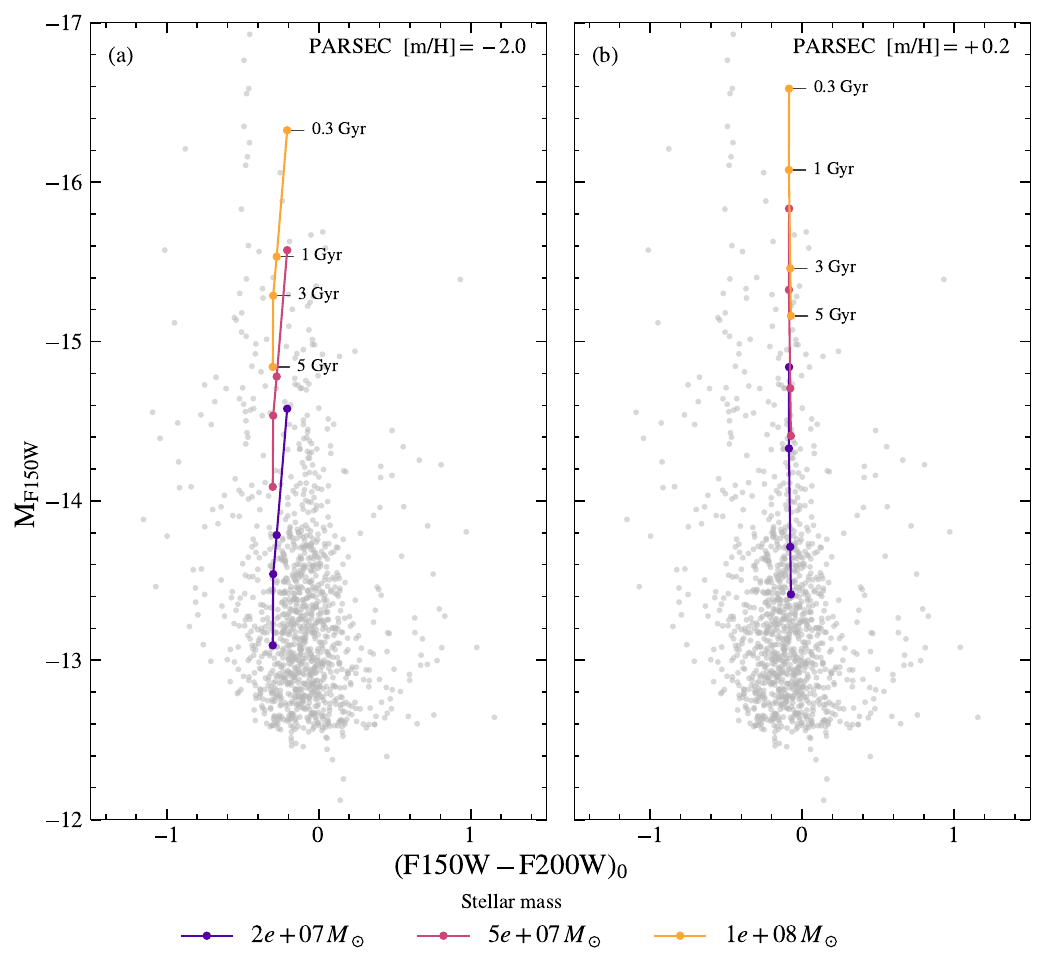}
    \caption{MACS~1149 compact source candidates compared with the PARSEC SSP models spanning stellar masses of $2\times10^7\,M_\odot,\ 5\times10^7\,M_\odot$, and $10^8\,M_\odot$ at ages of 0.3, 1, 3, and 5 Gyr. For each given mass, the age sequence increases from top to bottom. The left and right panels show models with $\rm [m/H]=-2.0$ and $+0.2$.}
    \label{fig:ssp_mass}
\end{figure}

The comparison in Figure \ref{fig:ssp} showed a fixed stellar mass of $2\times10^7\,M_\odot$ to illustrate the expected photometric evolution of massive GCs between the MACS~1149 and MACS~0416 epochs. Several of the most luminous MACS~1149 sources extend well above these fixed-mass sequences, raising the question of whether their luminosities require unusually large stellar masses. Alternatively, could these sources also be produced by younger stellar populations? Here we expand the range of SSP parameters to explore this question. 
Figure~\ref{fig:ssp_mass} compares the MACS~1149 population PARSEC SSP models spanning stellar masses of $2\times10^7\,M_\odot,\ 5\times10^7\,M_\odot$, and $10^8\,M_\odot$, and ages 0.3, 1, 3, and 5 Gyr, showing a metal-rich and metal-poor composition. The young ages are included to explore how stellar age and fading can affect the luminosities of the brightest compact sources, and should not be interpreted as age estimates for the MACS~1149 GCs. 

Once the age range is expanded, an interesting change in the model sequence becomes apparent. At a given age and metallicity, increasing stellar mass mainly increases the F150W luminosity while leaving the color unchanged, whereas changing age produces a more complex trajectory. This is especially noticeable for the metal-poor models, which become redder in $(\mathrm{F150W}-\mathrm{F200W})_0$ between 0.3 and $\sim1$~Gyr before turning back towards bluer colors at later ages. The behavior is seen in intermediate-age populations, where the near-infrared light can be strongly influenced by luminous evolved stars, particularly those in the AGB and TP-AGB phases \citep{Maraston2005}. Their contribution to the integrated light changes quickly with age and metallicity, making the near-infrared colors of intermediate-age populations especially sensitive to these relatively short-lived phases. The PARSEC+COLIBRI models used here explicitly follow the TP-AGB evolution through successive thermal-pulse cycles \citep{Marigo2013,Marigo2017}. By $3-5$ Gyr, the color evolution becomes notably weaker and begins to show vertical behavior seen in the older SSP sequences in Figure~\ref{fig:ssp}. The detailed shape of this intermediate-age evolution depends on the treatment of these evolved stellar phases and can therefore vary among stellar population models \citep{Conroy2013}.

This comparison helps show why luminosity alone cannot determine the nature of the brightest MACS~1149 sources. As stellar populations fade with age, an older, more massive population can occupy the same F150W magnitude space as a younger, less massive population, showing the degeneracy between the age and stellar mass in the observed CMD. Metallicity adds another uncertainty through its effect on the integrated color. These effects make it complicated to interpret the most luminous sources in particular. Although some of their elevated luminosities can be produced by younger stellar populations, the brightest objects extend into the regime of UCDs and stripped nuclei.

There is also a vertical sequence of sources near $(\mathrm{F150W}-\mathrm{F200W})\simeq-0.3$. Similar features have been observed in other JWST cluster fields, suggesting that foreground stars may contribute to this sequence. However, because of the overlap with SSP models, it is challenging to distinguish foreground stars from the compact stellar systems in MACS~1149.

The present photometry therefore cannot distinguish between these possibilities for individual sources. However, the models show that the luminous end of the MACS~1149 population should not be interpreted in terms of stellar mass alone.

\label{ssp_mass}

Taken together, the LF and stellar population comparisons place MACS~1149 within the emerging picture of GCS evolution now being revealed by \jwst. The elevated luminosities observed at this epoch are consistent with younger stellar populations at larger lookback times, while the comparison with MACS~0416 demonstrates the fading expected as these populations evolve toward the present day. In combination with the broad bright-side GCLF (Section~4.2), these results extend the growing observational baseline for tracing the luminosity distributions of GCSs across cosmic time.

\section{Conclusions}
\label{sec:conclusion}
In this study, we have used deep \jwst/NIRCam F115W, F150W, and F200W imaging of MACS~J1149.5+2223 ($z=0.544$), to identify and characterize its compact-source population and to place it within the emerging picture of GCS evolution at intermediate redshift. Our principal conclusions are:

\begin{enumerate}
    
    \item Sample and Completeness: PSF-fitting photometry and sharpness cut yield 1,202 unresolved compact-source candidates detected in all three filters. Artificial star tests give 80\% and 50\% completeness at $m_\mathrm{F150W}=28.80$ and 29.54 mag. Local background has a statistically significant but relatively small effect on source recovery, primarily in the cores and halos of bright cluster galaxies.
    
    \item Luminosity function: The completeness-corrected F150W LF of the 983 candidates in the primary cluster module rises steadily towards fainter magnitudes over the fitted range. The expected turnover ($m_0=33.66$ mag adopted from Abell~2744 and MACS~0417) lies more than 4 mag fainter than the faintest fitted bin included in the fit, so it's unobserved, and $m_0$ and $\sigma$ are degenerate. With $m_0$ fixed, the bright-side dispersion is $\sigma=1.71$ mag, and falls to 1.47 mag when the 67 brightest objects $(m_\mathrm{F150W}<28.0)$ are excluded. The fitted dispersion describes the bright-side slope only, is conditional on the adopted $m_0$, and is sensitive to the most luminous sources.
    
    \item Stellar-population fading: Fixed mass PARSEC SSP models pass through the main body of the observed color-magnitude distribution. The offset between MACS~1149 and MACS~0416 is in the direction expected from age fading. This supports age fading is an important contributor to the luminosity function evolution of GCSs, consistent with earlier JWST results. Age fading alone does not account for the brightest sources, which extend above the fixed-mass SSP sequences. These sources could be more massive GCs, UCDs, stripped nuclei, or Galactic foreground stars. However, we cannot separate the compact stellar systems on a source-by-source basis. 

    \item Degeneracies: Across the range of stellar masses, ages, and metallicities explored, age, mass, and metallicity are strongly degenerate at the luminous end. Photometry alone therefore cannot provide unique mass estimates or define a physically motivated luminosity boundary between ordinary GCs and more massive compact systems. 
    
\end{enumerate}

\section*{Acknowledgments}
J.M.B. gratefully acknowledges support from a University Graduate Fellowship provided by the School of Earth and Space Exploration at Arizona State University. The authors acknowledge the SKYSURF-IR collaboration for the production and development of the reduced \jwst/NIRCam mosaics used in this work.

\newpage

\section*{Data Availability}
The data presented in this work were obtained from the Mikulski Archive for Space Telescopes (MAST) at the Space Telescope Science Institute. The observations analyzed in this work are associated with JWST program 1176 at doi: 10.17909/xhha-f928.

\textit{Facility:} \jwst\ (NIRCam).

\textit{Software:} DAOPHOT \citep{Stetson1987}, IRAF
\citep{Tody1986,Tody1993}, Astropy
\citep{Astropy2013,Astropy2018,Astropy2022}, NumPy
\citep{Harris2020}, pandas \citep{McKinney2010}, Matplotlib
\citep{Hunter2007}, SciPy \citep{Virtanen2020}, statsmodels
\citep{Seabold2010}, and PARSEC v1.2S \citep{bressan2012}.
\bibliographystyle{aasjournal}
\bibliography{references}

\end{document}